\documentclass[a4paper,fleqn,usenatbib]{mnras}

\usepackage{newtxtext,newtxmath}

\usepackage[T1]{fontenc}
\usepackage{ae,aecompl}

\usepackage{graphicx}	
\usepackage{amsmath}	
\usepackage{amssymb}	
\usepackage{color} 

\title[The formation of giant planets in wide orbits]{The formation of giant planets in wide orbits by photoevaporation-synchronised migration }

\author[Guilera, Miller Bertolami \& Ronco]{O. M. Guilera,$^{1,2}$\thanks{E-mail: oguilera@fcaglp.unlp.edu.ar}
M. M. Miller Bertolami$^{1}$\thanks{E-mail: mmiller@fcaglp.unlp.edu.ar}
M. P. Ronco,$^{1,2}$
\\
$^{1}$Instituto de Astrof\'{\i}sica de La Plata, CCT La Plata, CONICET-UNLP, Paseo del Bosque S/N, B1900FWA La Plata, Argentina\\
$^{2}$Facultad de Ciencias Astron\'omicas y Geof\'{\i}sicas, UNLP, Paseo del Bosque S/N, B1900FWA La Plata, Argentina \\
}

\date{Accepted XXX. Received YYY; in original form ZZZ}

\pubyear{2017}

\begin{document}
\label{firstpage}
\pagerange{\pageref{firstpage}--\pageref{lastpage}}
\maketitle

\begin{abstract}
The discovery of giant planets in wide orbits represents a major challenge for 
planet formation theory. In the standard core accretion paradigm planets are expected to form at radial distances $\lesssim 20$~au in order to form massive cores (with masses $\gtrsim 10~\textrm{M}_{\oplus}$) able to trigger the gaseous runaway growth before the dissipation of the disc. This has encouraged authors to find modifications of the standard scenario as well as alternative theories like the formation of planets by gravitational instabilities in the disc to explain the existence of giant planets in wide orbits. However, there is not yet consensus on how these systems are formed. 

In this letter, we present a new natural mechanism for the formation of giant planets in wide orbits within the core accretion paradigm. If photoevaporation is considered, after a few Myr of viscous evolution a gap in the gaseous disc is opened. We found that, under particular circumstances planet migration becomes synchronised with the evolution of the gap, which results in an efficient outward planet migration. This mechanism is found to allow the formation of giant planets with masses $M_p\lesssim 1 M_{\rm Jup}$ in wide stable orbits as large as $\sim$130~au from the central star.
\end{abstract}



\section{Introduction}
The discovery of giant planets in very wide orbits by
\citet{2008Sci...322.1345K} and \citet{2008Sci...322.1348M} posed a challenge to
planet formation theory. In the core accretion mechanism giant planets
are expected to form at radial distances $a \lesssim 20$~au in order
to form cores which are massive enough to trigger the gaseous runaway
growth (with masses $M_p\gtrsim 10~\textrm{M}_{\oplus}$) before the
dissipation of the disc. While the core accretion mechanism was very
successful in explaining the properties of planets and planetary
systems discovered by radial velocities and transits, the explanation of giant
planets in wide orbits, like those surrounding Fomalhaut and HR 8799,
is difficult. \cite{2009ApJ...707...79D} performed simple numerical
experiments for the formation of planets in the core accretion
mechanism and concluded that even under the most favourable conditions
the formation of giant planets at $a \gtrsim 35$~au is not
possible. In addition, by performing N-body simulations
\cite{2009ApJ...707...79D} showed that planet-planet scattering was
not able to create stable wide-orbit systems such as HR 8799. They
concluded that massive (gas) giant planets in wide stable orbits with
$a \gtrsim 35$~au are formed by gravitational instabilities in the
disc. \cite{2011ApJ...731...74B} studied the formation of giant
planets in wide orbits by gravitational instabilities in discs of different masses  and around different possible
protostars, and concluded that the formation by gravitational
instabilities of gas giants with semi-major axes $a\simeq 30 \textendash 70$~au, eccentricities $e\simeq 0 \textendash 0.35$, and masses $M_p\simeq 1 \textendash 5$~$M_{\rm Jup}$ would be
possible if the disc fragments managed to survive to longer times.
Later, however, \cite{2013A&A...552A.129V} showed that disc fragments
have a low probability of survival due to inward migration of ejection
mechanisms during the embedded phase of star formation. Consequently,
\cite{2013A&A...552A.129V} concluded that the disc instability
mechanism was unable to form gas giants with semi-major axes $a
\lesssim $178 au and proposed that planets with semi-major axes from a
few tens to 150 au are likely to form via dynamical scattering or
other mechanisms. Therefore, the question of whether disc
instabilities can finally settle into stable wide orbits remains open.

Aside from the disc instability mechanism other alternatives have been
discussed like the recapture of free floating planets
\cite{2012ApJ...750...83P} or modifications within the core accretion
model. Examples of the latter are efficient outward type II migration
\citep{Crida-et-al-2009}, and high solid accretion rates
\citep{Lambrechts-et-al-2014}.  During the development of our previous
population synthesis work \citep{Ronco.et.al.2017} we serendipitously 
found that a small percentage of the total of the performed
simulations presented a giant planet in an extended orbit at the end
of the gaseous phase, between 40 au and 100 au. An inspection of the
results showed these protoplanets migrating outwards in synchrony with
the outer boundary of the density gap opened by photoionisation. In
this letter we present a detailed account of this phenomenon, which
results in a natural scenario for the formation of giant planets in
stable wide orbits within the framework of the core accretion paradigm
of planet formation.

The letter is organised as follows, in section \ref{modeling} we
briefly describe the model of planet formation with emphasis in the
upgrades with respect to our previous works. Then, in
section \ref{scenario_results} we describe the scenario and describe
our numerical results. Finally, on section \ref{conclusions} we make
some final statements and concluding remarks.

\section{Model and input physics}
\label{modeling}
 We adopt an improved version of the planet formation code
 P{\scriptsize LANETA}LP, adopted in our previous work
 \citep{Ronco.et.al.2017}. Specifically we have incorporated a
 standard 1D+1D treatment to solve the vertical and radial structure
 of a viscous irradiated accretion disc and incorporated the type I
 migration rates for non-isothermal discs. In what follows we briefly
 describe the implementation of these changes and the most relevant
 properties of the model for the present work.

\subsection{The gaseous disc}
\begin{figure*}
  \centering
  \includegraphics[angle= 0, width= 0.75\textwidth]{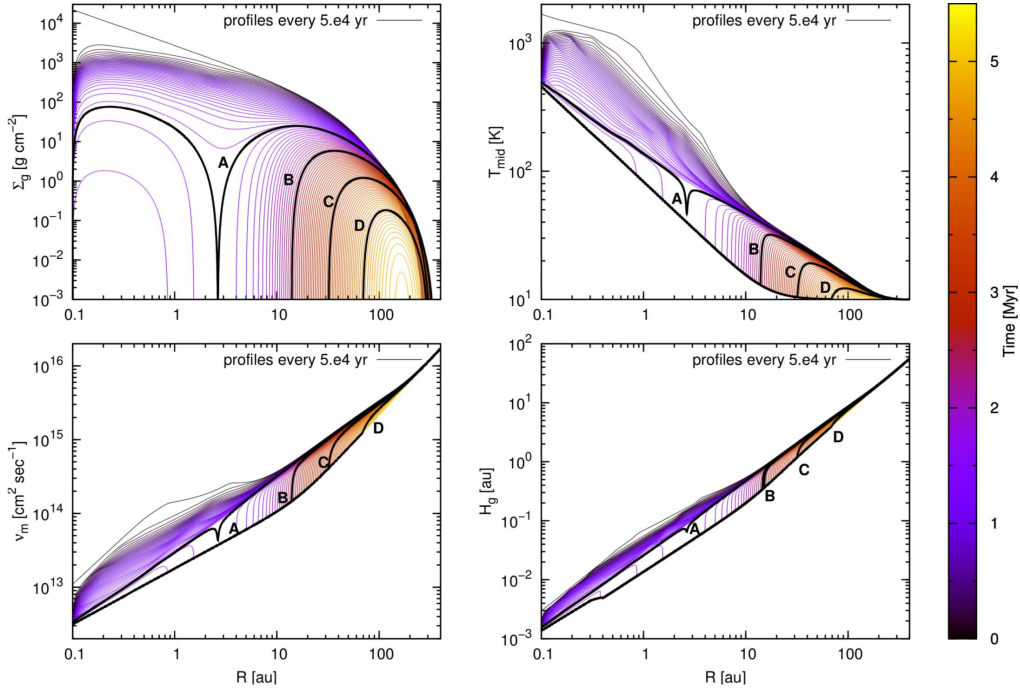}
  \caption{Time evolution of the radial profiles of the gas surface density ($\Sigma_{\text{g}}$; top left panel), the mid-plane temperature ($T_{\rm mid}$; top right panel), the mean viscosity ($\nu_{\rm m}$; bottom left panel), and the scale height of the disc ($H_{\rm g}$; bottom right panel). The radial profiles A,B,C, and D (black thick lines) represent the structure of the disc at $\sim 1.6$~Myr (when the gap is opened due to photoevaporation), at 2.5~Myr, at 3.5~Myr, and at 4.5~Myr, respectively. The disc is dissipated at $\sim 5$~Myr.}
  \label{fig:fig1}
\end{figure*}

We assume a thin, axisymmetric, irradiated disc in hydrostatic equilibrium. To solve the vertical structure of the disc we follow the classical methodology as in \citet{Papaloizou.Terquem.1999} and \citet{Alibert2005}, solving for each radial bin the following structure equations
\begin{equation}
  \frac{\partial P}{\partial z} = -\rho \Omega^2 z,\ \ \ \ 
  \frac{\partial F}{\partial z} = \frac{9}{4} \rho \nu \Omega^2,\ \ \ \  
  \frac{\partial T}{\partial z} = \nabla \frac{T}{P}\frac{\partial P}{\partial z}
 \label{eq:eq1} 
\end{equation}
where $P$, $\rho$, $F$, $T$ and $z$ represents the pressure, density,
radiative heat flux, temperature and vertical coordinate of the disc,
respectively. $\Omega$ is the Keplerian-frequency at a given radial
distance, while $\nu= \alpha c_s^2/\Omega$ is the viscosity
\citep{Shakura1973}, being $c_s^2= P/\rho$ the square of the locally
isothermal speed of the sound. We consider that heat is transported by
radiation and convection according to the standard Schwarzschild
criterion.  The actual temperature gradient $\nabla= d \log T / d \log
P$ in convective regions is computed with the help of the mixing
length theory, following the approach of
\cite{2012sse..book.....K}. On purely radiative regions the
temperature gradient ($\nabla_{\rm rad}$) is computed in the diffusive
approximation valid for optically thick discs
\begin{equation}
  \nabla_{\text{rad}}= \frac{3\kappa \rho F}{16 \sigma \Omega^2 z T^4},
  \label{eq:eq3} 
\end{equation}
where $\sigma$ is the Stefan-Boltzmann constant and $\kappa$ is the
local Rosseland mean opacity \citep{Bell.Lin.1994}. We adopt the
equation of state of an ideal diatomic gas $P= \rho k T / \mu
m_{\text{H}}$ where $k$ is the Boltzmann constant, $\mu= 2$ is the
mean molecular weight, $ m_{\text{H}}$ is the mass of the hydrogen
atom, and the the adiabatic gradient for a diatomic gas is
$\nabla_{\text{ad}}= 2/7$.

Following \citet{Papaloizou.Terquem.1999}, \citet{Alibert2005} and
\citet{Migaszewski2015} the boundary conditions at the surface of the
disc $H$, $P_s= P(z=H)$, $F_s= F(z=H)$, $T_s= T(z=H)$, and $F_0=
F(z=0)$, are given by
\begin{align}
  P_s &= \frac{\Omega^2 H \tau_{\text{ab}}}{\kappa_s},\ \ \ 
  F_s = \frac{3 \dot{M}_{\text{st}} \Omega^2}{8\pi} ,\ \ \  
  F_0 = 0, \nonumber \\  
  0 &= 2 \sigma \left( T_s^4 + T_{\text{irr}}^4 - T_{\text{b}}^4 \right) - \frac{9 \alpha \Omega k (T_s^4 + T_{\text{irr}}^4)^{1/4}}{8 \kappa_s \mu m_{\text{H}}} - F_s,
\label{eq:eq4} 
\end{align}
where $\tau_{\text{ab}}= 0.01$ is the optical depth, $T_{\text{b}}=
10$~K is the background temperature, and $\dot{M}_{\text{st}}$ is the
equilibrium accretion rate. The temperature associated with the
stellar irradiation is given by
\begin{equation}
  T_{\text{irr}}= T_{\star} \left[ \frac{2}{3\pi} \left( \frac{R_{\star}}{R} \right)^3 + \frac{1}{2} \left( \frac{R_{\star}}{R} \right)^2 \left( \frac{H}{R} \right) \left( \frac{d\log H}{d\log R} - 1 \right) \right]^{0.5},
  \label{eq:eq5} 
\end{equation}
where $R_{\star}= 2~R_{\odot}$ and $T_{\star}= 4000$~K are the radius
and effective temperature of the protostar, $R$ is the radial
coordinate, and $d\log H/d\log R= 9/7$.  To solve the vertical
structure of the disc we follow a numerical approach similar to the
one described in \citet{Alibert2005} and \citet{Migaszewski2015}.
This allows us to obtain the relation $\nu_{\text{m}}(\Sigma_{\text{g}}, R)$, 
where $\Sigma_{\text{g}}$ is the gas surface density,
required to solve the radial evolution of the disc (eq. \ref{eq:eq6})
as well as other relationships which are needed to compute the planet
migration rates (see next section).


The time evolution of the gas surface density is represented by a
diffusion equation \citep{Pringle1981}
\begin{equation}
  \frac{\partial \Sigma_{\text{g}}}{\partial t}= \frac{3}{R}\frac{\partial}{\partial R} \left[ R^{1/2} \frac{\partial}{\partial R} \left( \nu_{\text{m}} \Sigma_{\text{g}} R^{1/2}  \right) \right] + \dot{\Sigma}_{\text{w}}(R), 
\label{eq:eq6}
\end{equation}
where $\dot{\Sigma}_{\text{w}}(R)$ represents a sink term due to the
photoevaporation by the central star, which is given by
\citep{DAngeloMarzari2012}
\begin{equation}
  \dot{\Sigma}_{\text{w}}(R) = 
  \begin{cases}
    \dot{\Sigma}^{g}_{\text{w}}\exp{\left[\frac{1}{2}\left(1-\frac{R_{\text{g}}}{R}\right)\right]}\left(\dfrac{R_{\text{g}}}{R}\right)^2 & \text{ if $R \le R_{\text{g}}$},  \\
    \\
    \dot{\Sigma}^{g}_{\text{w}}\left(\frac{R_{\text{g}}}{R}\right)^{5/2} & \text{ if $R > R_{\text{g}}$},
  \end{cases} 
\label{eq:eq7}
\end{equation}
where $R_{\text{g}}$ is the distance along the midplane beyond which gas at the disk surface is unbound. We take $R_{\text{g}}=10$~au as it is usually assumed for solar-mass stars. The photoevaporation rate at $R_{\text{g}}$ is
represented by $\dot{\Sigma}^{g}_{\text{w}}$ and is given by
\begin{equation}
  \dot{\Sigma}^{g}_{\text{w}} = 1.16 \times 10^{-11}\sqrt{f_{41}}\left(\frac{R_{\text{g}}}{1\text{au}}\right)^{-1.5}\frac{M_\odot}{\text{au}^{2}\text{yr}},
  \label{eq:eq8}
\end{equation}
where $f_{41}$ is the rate of EUV ionising photons emitted by the star in units of $10^{41}$~sec$^{-1}$.

\subsection{Planet growth and planet migration}
We considered the formation of a planet in the standard core accretion
mechanism by the concurrent accretion of solid and gas. Initially, the
core of the planet grows by the accretion of planetesimals in the
oligarchic growth regime. As the planet grows, it binds the
surrounding gas, reaches a critical mass (of about a tens Earth
masses) and accretes a substantial gaseous envelope.  The gas
accretion rate is estimated from the results of
\citet{Guilera2010,Guilera2014} following the prescriptions of
\citet{Ida-Lin2004} and \citet{Miguel2011}. We include the limitation
in the gas accretion rate due to the capability of the disc to supply
enough material \citep{Mordasini2009}, and the possibility that the
planet opens a gap in the gaseous disc \citep{TanigawaIkoma2007}. The
population of planetesimals is considered to evolve by the drift of
planetesimals due to the nebular gas including the Epstein, Stokes and
quadratic regimes, and the accretion and ejection by the embryos
---see \citet{ Ronco.et.al.2017} for details. We consider that the
evolution of the eccentricities and inclinations of the planetesimal
population are due to two main processes, the embryo gravitational
excitations and the damping due to the nebular gas.

Of special interest for the present work is the treatment of planet
migration.  The interaction between the gaseous disc and the planet
leads to an exchange of angular momentum producing torques that cause
the planet's migration along the disc. If the planet is not able to
open a gap in the disc we consider that it migrates under type I
migration
\begin{equation}
 \left(\frac{da_{\text{P}}}{dt}\right)_{\text{migI}} = -2 a_{\text{P}}\frac{\Gamma}{L_{\text{P}}},
 \label{eq:eq9}
\end{equation}
where $a_{\text{P}}$ is the semi-major axis of the planet,
$L_{\text{P}}$ is the angular momentum of the planet, and $\Gamma$ is
the total torque. In this work, we incorporate the prescriptions of
\citet{paardekooper_etal2010, paardekooper_etal2011} to calculate the
total torque for non-isothermal discs.  When the planet is massive
enough to open a gap in the disc \citep{Crida2006}, the migration of
the planet switches to the known type II migration \citep{Armitage2007, 
Mordasini2009}. 


\section{Results and Discussion}
\label{scenario_results}

\begin{figure*}
  \centering
  \includegraphics[angle= 0, width= 0.75\textwidth]{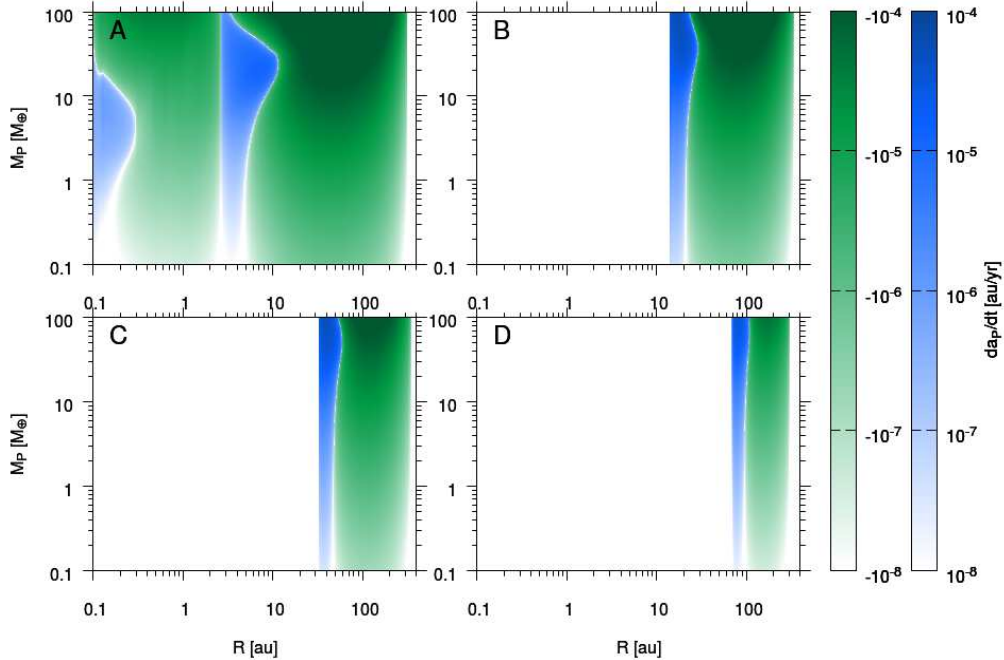}
  \caption{Migration rates for the four disc structures (A,B, C and D) shown in Fig. \ref{fig:fig1} as a function of the distance from the central star and mass of the planet at different times of the disc evolution. Panel A represents the migration map when the gap is opened in the disc. Panels B, C, and D represent the migration map at 2.5~Myr, at 3.5~Myr, and at 4.5~Myr, respectively. The blue palette represents the outward migration zone, while the green palette represents the inward migration zones.}
  \label{fig:fig2}
\end{figure*}

\begin{figure}
  \centering
  \includegraphics[angle= 0, width= 0.4\textwidth]{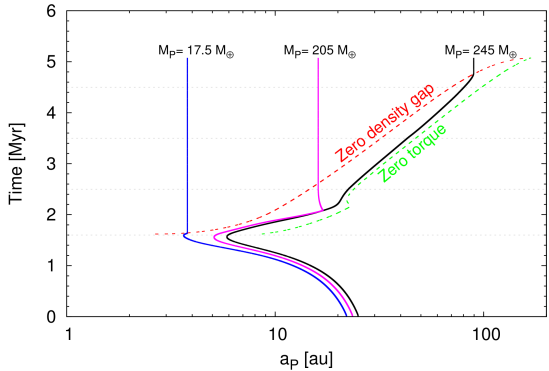}
  \caption{Evolution of the planet semi-major axis as a function of time 
    for our reference case.
    The red dashed line indicates the evolution of the outer
    boundary of the photoevaporation gap and the green dashed line the
    evolution of the zero torque location in the outer part of the
    disc. Grey horizontal lines correspond to times A, B, C and D in
    Figs. \ref{fig:fig1} and \ref{fig:fig2} (from bottom to
    top). Different curves corresponds to protoplanets that entered
    the ROM with different masses and had different evolutions (see
    text).}
  \label{fig:fig3}
\end{figure}

 Fig. \ref{fig:fig1} shows the evolution of gas density
 $\Sigma_{\text{g}}$, mid-plane temperature $T_{\rm mid}$, mean
 viscosity $\nu_{\text{m}}$ and gas-disc scale-height $H_{\text{g}}$
 for our reference model. This model was computed for a disc with
 a mass of $M_{\rm d}=0.1 M_\odot$, a viscosity parameter
 $\alpha=10^{-3}$, a characteristic radius of $R_c= 39$~au, and an
 initial surface density profile of $\Sigma_{\text{g}}(R)=\Sigma_0
 (R/R_c)^{-\gamma} \exp((R/Rc)^{-(2-\gamma)})$ with $\gamma=0.9$
 \citep{Andrews2010}, surrounding a pre-main sequence star of $1\,
 M_\odot$. As shown in Fig. \ref{fig:fig1} (curve A) photoevaporation
 opens a gap in the disc at $R\simeq 2.65$~au after $\sim 1.6$~Myr of
 evolution. Due to the strong dependence of type I migration rates
 with the slope of the density and temperature profiles
 \citep{paardekooper_etal2011} the opening of the gap leads to the
 creation of a region of outward type I migration (ROM; blue zones in
 Fig. \ref{fig:fig2}). As time passes, the combined effects of viscous
 evolution and photoevaporation widen the gap and push the ROM
 outwards (Fig. \ref{fig:fig1}, curves B, C and D ). The ROM is
 bounded at any time by two rings, the zero density inner boundary and
 the zero torque outer boundary. Note that planets migrating according
 to type I migration cannot cross the zero torque ring at any moment.
 Within the ROM, if a planet is massive enough to have an efficient
 type I migration then it will move outwards faster than the inner
 boundary does but, as it cannot cross the outer boundary it will be
 forced to move inside the ROM. As long as the disc is not dissipated
 the mass of the planet can only increase, which in most cases means
 more efficient migration preventing the planet to lag behind the
 ROM. It is possible to imagine that a forming planet
 could be trapped in the ROM and grow as it migrates outwards
 synchronised with the outwards evolution of the ROM. We term these
 objects synchronised outwards migrating planets (SOMP). SOMPs will
 keep evolving outwards until the gas disc is dissipated or
 they become massive enough to change to a type II migration regime
 \citep{Crida2006}.

Fig. \ref{fig:fig3} shows the evolution of one of this SOMPs in our
reference disc model (black curve, Fig. \ref{fig:fig3}). The embryo,
having initially a mass of $0.01~M_{\oplus}$ and located at 25~au,
migrates inwards with increasing speed as the mass of the protoplanet
increases by the accretion of planetesimals of $\sim 10$~km of radius.  
By the time the gap is opened at $\sim 1.6$~Myr the
forming planet already has a mass of $\sim 10~M_{\oplus}$ and
undergoes efficient type I migration. As a consequence, as
photoevaporation opens the gap, and changes the slopes of the
temperature and density profiles, the direction of migration changes
suddenly from a normal inward type I migration into an outward
migration. After that point the planet evolves outward inside the ROM
(bounded by the red and green curves in Fig. \ref{fig:fig3}). The
outward migration is finished at $\sim 4.8$~Myr, when the gas
disc becomes too thin. By the time migration stops the planet has
increased its mass up to $M_p \sim 250~M_{\oplus}$ and reached a wide
stable orbit at $\sim 90$~au. This example shows that
photoevaporation coupled to type I migration can naturally lead to the
formation of massive planets in wide orbits within the core accretion
framework. For a planet to become synchronised with the ROM and to evolve
to very wide orbits the planet needs to reach the ROM with enough mass
to have an efficient type I migration. If the mass is too small,
outwards migration will be inefficient and the planet will be caught
up by the zero density gap, stopping its gas accretion and migration
(blue curve, Fig. \ref{fig:fig3}).  On the contrary, if
the planet is too massive by the time it enters the ROM it might only
evolves for a short time before becoming massive enough to open a gap
and shift to a type II migration regime (pink curve, Fig. \ref{fig:fig3}). 
Consequently, for the SOMP phenomenon to develop
and produce planets in very wide orbit, the protoplanet needs to enter the
ROM with masses in a narrow range around 10~$M_\oplus$.

\begin{figure}
  \centering
  \includegraphics[angle= 0, width= 0.4\textwidth]{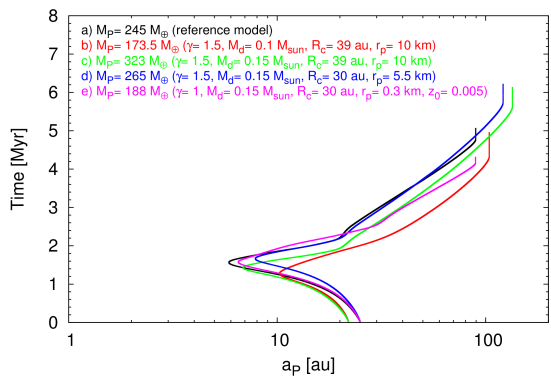}
  \caption{Evolution of the planet semi-major axes as a function of time 
    for different SOMPs formed in different discs. $\text{M}_{\text{P}}$ represents the final masses of the planets. Quantities between brackets the parameters of the different discs. Specially interest remains in case e) where the disc have a metalicity similar to HR8799.}
  \label{fig:fig4}
\end{figure}

 In Fig.~\ref{fig:fig4} we show similar cases of
 photoevaporation-synchronised migration computed under different
 assumptions for the structure of the disc and the photoevaporation
 rates (for all cases we use $\alpha= 10^{-3}$). We check the stability of the discs adopting the Toomre $Q$ parameter \citep{Toomre1964}.  
 As can be seen in
 Fig.~\ref{fig:fig4} the photoevaporation-synchronised scenario is
 able to predict the formation of planets of $0.55\textendash1~M_{\rm
   Jup}$ as far as $90\textendash130$~au from the central star within
 the core accretion framework.  In these exploratory simulations we find that
 most of our giant planets are formed with a high content of solids,
 and consequently should have smaller radii than planets formed
 through the disc instability mechanism. This opens the possibility to
 test both scenarios \citep{Spiegel-Burrows2012}. 
While the quantitative results predicted by this scenario are dependent on the disc mass, and the adopted photoevaporation \citep{Gorti2009, Owen2012} and type I migration models, the qualitative behaviour predicted here should be present in all models with an outward moving gap and  a type I migration model dependent on the disc density profile. 

%

\section{Conclusions}
\label{conclusions}
 We have presented a new scenario for the formation of giant planets
 in wide orbits within the core accretion paradigm that arises
 naturally from the synchronization of planet migration and disc
 photoevaporation. We have shown that this scenario is able to form
 Jupiter like-planets in wide orbits as large as $\sim$130~au.  The
 scenario requires the mass of the protoplanet to be fine tuned at the
 moment it is trapped in the region of outwards type I migration, but
 our computations show that this scenario works in a wide region of
 the parameter space (Fig.~\ref{fig:fig4}).  The fact that discs
   with large inner holes are more common around Herbig Ae/Be
   than around solar mass stars \citep{Maaskant2013} suggests that
   this mechanism might be more viable to form giant planets in wide orbits around A-type stars
   \citep{2008Sci...322.1345K,2008Sci...322.1348M}. Further
 exploration of this scenario by means of population synthesis studies
 is required to assess the fraction of protoplanets that
 undergo photoevaporation synchronised migration and, also, to
 establish the general properties (masses, radii and semi-major axes)
 predicted by this scenario. In particular, it remains to be seen if
 multiple planet systems like those observed in HR8799 can be formed
 within this picture once resonances are taken into account. We plan
 to explore these possibilities in future works.

\section*{Acknowledgements}

We thank Yann Alibert and the anonymous referee for their suggestions to improve this work. O.M.G and M.P.R are supported by CONICET and UNLP through grants PIP-0436 and G-144. M3B is partially supported by ANPCyT through grant PICT-2014-2708 and
by a Return Fellowship from the Alexander von Humboldt Foundation.




\bibliographystyle{mnras}
\bibliography{biblio} 








\bsp	
\label{lastpage}
\end{document}